\documentclass[prd,aps,preprintnumbers,nofootinbib,superscriptaddress,notitlepage,floatfix,11pt]{revtex4}
\usepackage{stmaryrd}
\usepackage{epsfig}
\usepackage{amsfonts}
\usepackage{amsmath}
\usepackage{graphicx}
\usepackage{color}
\usepackage{amssymb,amsmath,psfrag,slashed,graphicx}
\usepackage[normalem]{ulem}

\def \nn {\nonumber}

\def\red{\textcolor[rgb]{1.00,0.00,0.00}}

\allowdisplaybreaks

\begin{document}
\title{Deciphering Weak Decays of Triply Heavy Baryons by SU(3) Analysis}

\affiliation{School of Physics and Microelectronics, Zhengzhou University, Zhengzhou, Henan 450001, China}
\affiliation{INPAC, Key Laboratory for Particle Astrophysics and Cosmology (MOE),
Shanghai Key Laboratory for Particle Physics and Cosmology,
School of Physics and Astronomy, Shanghai Jiao Tong University, Shanghai 200240, China}

\author{Fei Huang~\footnote{fhuang@sjtu.edu.cn}}
\affiliation{INPAC, Key Laboratory for Particle Astrophysics and Cosmology (MOE),
Shanghai Key Laboratory for Particle Physics and Cosmology,
School of Physics and Astronomy, Shanghai Jiao Tong University, Shanghai 200240, China}

\author{Ji Xu~\footnote{Corresponding author. xuji\_phy@zzu.edu.cn}}
\affiliation{School of Physics and Microelectronics, Zhengzhou University, Zhengzhou, Henan 450001, China}

\author{Xi-Ruo Zhang}
\affiliation{School of Physics and Microelectronics, Zhengzhou University, Zhengzhou, Henan 450001, China}

\begin{abstract}
Baryons with three heavy quarks are the last missing pieces of the lowest-lying baryon multiplets in the quark model after the discovery of doubly heavy baryons. In this work, we study nonleptonic weak decays of triply heavy baryons $\Omega_{ccc}^{++}$, $\Omega_{bbb}^{-}$, $\Omega_{ccb}^{+}$, and $\Omega_{cbb}^{0}$. Decay amplitudes for various processes have been parametrized in terms of the SU(3) irreducible nonperturbative amplitudes. A number of relations for the partial decay widths can be deduced from these results that can be examined in future. Some decay channels and cascade decay modes which likely to be used to reconstruct the triply heavy baryons have been also listed.
\end{abstract}

\maketitle

\section{Introduction}
Triply heavy baryons which consist of three heavy $c$ or $b$ quarks are of great theoretical interests since they refrain from light quark contaminations. Being baryonic analogues of heavy quarkonium, the study of triply heavy baryons can help us to better understand the dynamics of strong interactions and would yield sharp tests for QCD. Besides, these baryons also provide particular information on the three body static potential. Previous studies on triply heavy baryons mainly concentrated on spectroscopy, relevant theoretical tools such as nonrelativistic constituent quark model (NRCQM)~\cite{Vijande:2004at}, potential nonrelativistic QCD (pNRQCD)~\cite{Brambilla:2005yk}, and the QCD sum rule (QCDSR)~\cite{Zhang:2009re} have been developed to investigate the nature of these baryons.

\begin{figure}
\includegraphics[width=0.65\columnwidth]{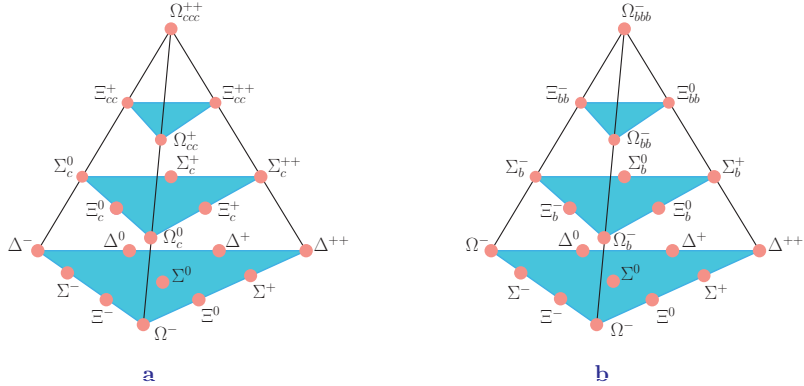}
\caption{Baryons with spin $3/2$ made from four quarks of the types $u,d,s,c$ (a) and $u,d,s,b$ (b). Triply heavy baryons $\Omega_{ccc}^{++}$ and $\Omega_{bbb}^{-}$ are localized in the highest layers.}
\label{omegacccandbbb}
\end{figure}

In the past decades, hadron spectroscopy has experienced a continuous progress. Since 2016, the BESIII Collaboration has reanalyzed the singly charmed baryon decays with higher precision \cite{ParticleDataGroup:2016lqr,BESIII:2015bjk}. One milestone for the doubly charmed baryon spectroscopy is the discovery of $\Xi_{cc}^{++}$ by the LHCb Collaboration~\cite{Aaij:2017ueg,Aaij:2018gfl}. Afterwards, baryons with three heavy quarks are the last missing pieces of the lowest-lying baryon multiplets in quark model, with this in mind, it is timely and meaningful to analyze triply heavy baryon on both theoretical and experimental sides at this stage. The flavor SU(5) group includes all types of baryons containing zero, one, two or three heavy quarks. One should note that the differences of quark masses break the flavor symmetry, the larger the group, the bigger the amount of breaking. The masses of three light quarks $u, d, s$ are much smaller than the masses of $c$ and $b$ quarks, this makes the flavor SU(3) symmetry basically maintained in weak decays of heavy baryons. However, the SU(5) group algebra helps us identify the triply heavy baryons which are interested in this paper. As an example, two sets of corresponding baryons projected along the $u,d,s,c$ and $u,d,s,b$ quarks are depicted in Fig.~\ref{omegacccandbbb}.

The production of triply heavy baryons is difficult and no experimental signal for any of them has been observed yet. The production rate of triply heavy charmed baryon in $e^+ e^-$ collision has been estimated to be very small~\cite{Baranov:2004er}, however, a recent investigation finds that around $10^{4}-10^{5}$ events of triply heavy baryons can be accumulated for $10\,fb^{-1}$ integrated luminosity at LHC~\cite{Chen:2011mb}. The heavy quarks can be produced via $gg$ fusion and quark-antiquark annihilation at hadron colliders. LHC and the future high luminosity LHC provide us a good chance to discover these triply heavy baryons. LHC has helped us find out doubly heavy baryon, undoubtedly, it will provide a sustained progress in heavy baryon field as well as the breadth and depth necessary for a vibrant research
environment.

Many studies about the triply heavy baryons can be found in the literature~\cite{Jia:2006gw,Meinel:2010pw,Wang:2011ae,Aliev:2012tt,Brown:2014ena,Wei:2015gsa,Thakkar:2016sog,Chen:2011mb,GomshiNobary:2004mq,
Flynn:2011gf,Geng:2017mxn,Liu:2019vtx,Yang:2019lsg}, however, despite the great progress, little attention has been paid to the decay properties. Various types of weak decays of triply heavy baryons occur, but unfortunately, a universal dynamical (factorization) approach has not be established yet. There are several distinct energy scales involved in the weak decays of triply heavy baryons which make the systematic factorization unavailable at present, these are the mass $m$ of heavy $c$ or $b$ quark, the momentum of the heavy quark $mv$, the off-shell energy of the heavy quark $m v^2$, and the energy of light hadron in the final state. This poses an obstacle for us to predict the decay width of triply heavy baryons. On the other hand, the approach of flavor SU(3) symmetry allows us to relate decay modes in the $b$ and $c$-hadron decays despite the unknown non-perturbative dynamics of QCD~\cite{Savage:1989ub,Gronau:1994rj,Grinstein:1996us,He:1998rq,Deshpande:2000jp,Deshpande:1994ii,Chiang:2003pm,Chiang:2006ih,
Li:2007bh,Wang:2008rk,Chiang:2008zb,Cheng:2014rfa,He:2014xha,He:2015fwa,He:2015fsa,Lu:2016ogy,Cheng:2016ejf,Cheng:2012xb,Li:2012cfa,Li:2013xsa,
Wang:2017azm,Shi:2017dto,Wang:2018utj,Li:2021rfj}. In this work, we consider nonleptonic decay channels of triply heavy baryons by utilizing flavor SU(3) analysis, it is an extension and supplement of a series of previous works. Ref.~\cite{Wang:2018utj} has discussed semileptonic and nonleptonic decay modes of $\Omega_{ccc}^{++}$, $\Omega_{bbb}^{-}$, $\Omega_{ccb}^{+}$ and $\Omega_{bbc}^{0}$. Beyond these modes, some two-body and three-body decay which are not covered in previous work would be considered in this work. Some particular interesting signature modes of triply heavy baryon decays such as $\Omega_{ccc}^{++} \to \Omega_{sss}^{-} + 3\pi^+$ will be discussed in the frame of SU(3) analysis. The main motivation of this work is to provide some suggestions which may help experimentalists find triply heavy baryons in future.

The present manuscript is arranged as follows. In Sec.~\ref{sec:particle_multiplet}, we present the irreducible forms of baryon and meson states under flavor SU(3) symmetry. In Sec.~\ref{sec:nonleptonic}, nonleptonic decays of triply charmed baryon $\Omega_{ccc}^{++}$, triply bottom baryon $\Omega_{bbb}^{-}$, the mixed triply heavy baryons $\Omega_{ccb}^{+}$ and $\Omega_{bbc}^{0}$ will be studied in order. A
short summary is given in the last section.

\section{Particle Multiplets}
\label{sec:particle_multiplet}
In this section, we will collect the representations for hadron multiplets under the flavor SU(3) group. The best determination of the magnitudes of CKM matrix elements and the Cabibbo parametrization will be also presented.

We start with the baryon sector. The initial triply heavy baryon singlet is given by
\begin{eqnarray}
  \left(\Omega_{ccc}^{++}\right)\,,\quad \left(\Omega_{bbb}^{-}\right)\,,\quad  \left(\Omega_{ccb}^{+}\right)\,,\quad \left(\Omega_{bbc}^{0}\right)\,.
\end{eqnarray}
The doubly heavy baryons are an SU(3) triplet:
\begin{eqnarray}
  \begin{aligned}
T_{c c}=\left(\begin{array}{c}
\Xi_{c c}^{++} \\
\Xi_{c c}^{+} \\
\Omega_{c c}^{+}
\end{array}\right)\,, \quad T_{b c}=\left(\begin{array}{c}
\Xi_{b c}^{+} \\
\Xi_{b c}^{0} \\
\Omega_{b c}^{0}
\end{array}\right)\,, \quad T_{b b}=\left(\begin{array}{l}
\Xi_{b b}^{0} \\
\Xi_{b b}^{-} \\
\Omega_{b b}^{-}
\end{array}\right) .
\end{aligned}
\end{eqnarray}
Singly charmed baryons with two light quarks can form an antitriplet or sextet which are
\begin{eqnarray}
 T_{\bf{c\bar 3}}= \left(\begin{array}{ccc} 0 & \Lambda_c^+  &  \Xi_c^+  \\ -\Lambda_c^+ & 0 & \Xi_c^0 \\ -\Xi_c^+   &  -\Xi_c^0  & 0
  \end{array} \right)\,, \quad T_{\bf{c 6}}=\left(\begin{array}{ccc}
\Sigma_{c}^{++} & \frac{1}{\sqrt{2}} \Sigma_{c}^{+} & \frac{1}{\sqrt{2}} \Xi_{c}^{\prime+} \\
\frac{1}{\sqrt{2}} \Sigma_{c}^{+} & \Sigma_{c}^{0} & \frac{1}{\sqrt{2}} \Xi_{c}^{\prime 0} \\
\frac{1}{\sqrt{2}} \Xi_{c}^{\prime+} & \frac{1}{\sqrt{2}} \Xi_{c}^{\prime 0} & \Omega_{c}^{0}
\end{array}\right) \,.
\end{eqnarray}
Light baryons made of three light quarks can group into an SU(3) octet and a decuplet. The octet has the expression:
\begin{eqnarray}
T_8= \left(\begin{array}{ccc} \frac{1}{\sqrt{2}}\Sigma^0+\frac{1}{\sqrt{6}}\Lambda^0 & \Sigma^+  &  p  \\ \Sigma^-  &  -\frac{1}{\sqrt{2}}\Sigma^0+\frac{1}{\sqrt{6}}\Lambda^0 & n \\ \Xi^-   & \Xi^0  & -\sqrt{\frac{2}{3}}\Lambda^0
  \end{array} \right) \,.
\end{eqnarray}
The indices of the decuplet is symmetric, it can be written in a compact form,
\begin{eqnarray}
 T_{10} = \frac{1}{\sqrt{3}}\left(\left(\begin{array}{ccc}
\sqrt{3} \Delta^{++} & \Delta^{+} & \Sigma^{\prime+} \\
\Delta^{+} & \Delta^{0} & \frac{\Sigma^{\prime 0}}{\sqrt{2}} \\
\Sigma^{\prime+} & \frac{\Sigma^{\prime 0}}{\sqrt{2}} & \Xi^{\prime 0}
\end{array}\right),\left(\begin{array}{ccc}
\Delta^{+} & \Delta^{0} & \frac{\Sigma^{\prime 0}}{\sqrt{2}} \\
\Delta^{0} & \sqrt{3} \Delta^{-} & \Sigma^{\prime-} \\
\frac{\Sigma^{\prime 0}}{\sqrt{2}} & \Sigma^{\prime-} & \Xi^{\prime-}
\end{array}\right),\left(\begin{array}{ccc}
\Sigma^{\prime+} & \frac{\Sigma^{\prime 0}}{\sqrt{2}} & \Xi^{\prime 0} \\
\frac{\Sigma^{\prime 0}}{\sqrt{2}} & \Sigma^{\prime-} & \Xi^{\prime-} \\
\Xi^{\prime 0} & \Xi^{\prime-} & \sqrt{3} \Omega^{-}
\end{array}\right)\right) \,.
\end{eqnarray}

For the meson sector, the light pseudoscalar mesons form an octet:
\begin{eqnarray}
 M_{8}=\begin{pmatrix}
 \frac{\pi^0}{\sqrt{2}}+\frac{\eta_8}{\sqrt{6}}  &\pi^+ & K^+\\
 \pi^-&-\frac{\pi^0}{\sqrt{2}}+\frac{\eta}{\sqrt{6}}&{K^0}\\
 K^-&\bar K^0 &-2\frac{\eta_8}{\sqrt{6}}
 \end{pmatrix} \,.
\end{eqnarray}
Please note that the $\eta$ in our calculations is only considered as a member of octet, while the singlet $\eta_1$ is not considered here to avoid the octet-singlet mixture complexity. The charmed and bottom mesons form similar SU(3) antitriplet,
\begin{eqnarray}
D_i=\left(\begin{array}{ccc} D^0, & D^+, & D^+_s  \end{array} \right) \,, \quad  B_i=\left(\begin{array}{ccc} B^-, & \overline{B}^0, & \overline{B}^0_s  \end{array} \right) \,.
\end{eqnarray}

Here we also present the best determination of the magnitudes of the CKM matrix elements~\cite{ParticleDataGroup:2020ssz}
\begin{eqnarray}\label{CKMNum}
  \left[\begin{array}{lll}
\left|V_{u d}\right| & \left|V_{u s}\right| & \left|V_{u b}\right| \\
\left|V_{c d}\right| & \left|V_{c s}\right| & \left|V_{c b}\right| \\
\left|V_{t d}\right| & \left|V_{t s}\right| & \left|V_{t b}\right|
\end{array}\right]=\left[\begin{array}{ccc}
0.97370 \pm 0.00014 & 0.2245 \pm 0.0008 & 0.00382 \pm 0.00024 \\
0.221 \pm 0.004 & 0.987 \pm 0.011 & 0.0410 \pm 0.0014 \\
0.0080 \pm 0.0003 & 0.0388 \pm 0.0011 & 1.013 \pm 0.030
\end{array}\right] \,,
\end{eqnarray}
and the Cabibbo parametrization formalism
\begin{eqnarray}
  \left[\begin{array}{ll}
V_{u d} & V_{u s} \\
V_{c d} & V_{c s}
\end{array}\right] = \left[\begin{array}{cc}
\cos \theta_{c} & \sin \theta_{c} \\
-\sin \theta_{c} & \cos \theta_{c}
\end{array}\right] \,,
\end{eqnarray}
to make the following discussions more comprehensible.

To depict the processes of various decay modes under the frame of flavor SU(3) analysis, we need to construct the hadron-level effective Hamiltonian in addition to the representations for initial and final states which have been listed above.  It is necessary to point out that a hadron in the final state must be created by its antiparticle field. For instance, for a $\Sigma_{c}^{++}$ appearing in the final state, we need the $\overline{\Sigma_{c}^{++}}$ in the Hamiltonian. The construction of hadron-level effective Hamiltonian will shown in the next section.

\section{Nonleptonic decays of triply heavy baryons}
\label{sec:nonleptonic}
\subsection{Nonleptonic $\Omega_{ccc}^{++}$ decays}
We start with the nonleptonic $\Omega_{ccc}^{++}$ decays. We have neglected penguin contributions in charm-quark decays since they are highly suppressed by the relevant CKM matrix elements. Tree operators of charm-quark decays into light quarks are categorized into three groups: Cabibbo-allowed, singly Cabibbo-suppressed, and doubly Cabibbo-suppressed,
\begin{eqnarray}
 c\to s \bar d u \,,  \;\;\; c\to u \bar dd/\bar ss \,, \;\;\; c\to  d \bar s u \,.
\end{eqnarray}
These tree operators transform under the flavor SU(3) symmetry as ${\bf  3}\otimes {\bf\bar 3}\otimes {\bf3}={\bf  3}\oplus {\bf  3}\oplus {\bf\bar 6}\oplus {\bf {15}}$. Thus the effective Hamiltonian can be decomposed in terms of a vector $H_3$; a traceless tensor antisymmetric in upper indices, $H_{\overline6}$; a traceless tensor symmetric in upper indices, $H_{15}$. The representation $H_3$ will vanish as an approximation by taking $V_{cd}^*V_{ud} = -V_{cs}^*V_{us}\simeq -\sin\theta_{c}$~\cite{Wang:2017azm}. The nonzero components of hadron-level Hamiltonian are listed below:
\begin{eqnarray}
(H_{\overline 6})^{31}_2=-(H_{\overline 6})^{13}_2=1 \,,\;\;\;
 (H_{15})^{31}_2= (H_{15})^{13}_2=1 \,, \qquad&&\rm{Cabibbo~allowed \,;} \nn\\
 (H_{\overline 6})^{31}_3 =-(H_{\overline 6})^{13}_3 =(H_{\overline 6})^{12}_2 =-(H_{\overline 6})^{21}_2 =\sin\theta_{c}\,, \qquad&&\rm{Singly~Cabibbo~suppressed \,;} \nn\\
 (H_{15})^{31}_3= (H_{15})^{13}_3=-(H_{15})^{12}_2=-(H_{15})^{21}_2= \sin\theta_{c}\,,  \qquad&&\rm{Singly~Cabibbo~suppressed \,;} \nn\\
 (H_{\overline 6})^{21}_3=-(H_{\overline 6})^{12}_3=\sin^2\theta_{c},\;\;
 (H_{15})^{21}_3= (H_{15})^{12}_3=\sin^2\theta_{c} \,, \qquad&&\rm{Doubly~Cabibbo~suppressed \,.} \nn\\
\end{eqnarray}

\begin{figure}
\includegraphics[width=0.30\columnwidth]{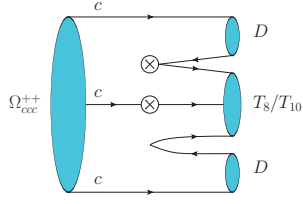}
\caption{The Feynman diagrams for $\Omega_{ccc}^{++}$ decays into two $D$-mesons and a light baryon (octet or decuplet).}
\label{1.1DDT810}
\end{figure}

For $\Omega_{ccc}^{++}$ decays into two $D$-mesons and a light baryon, the corresponding Hamiltonian can be constructed as
\begin{eqnarray}\label{cccA}
	{\cal H}_{\textit{eff}}&=& a_1  \Omega_{ccc}^{++} \varepsilon_{ijk} (\overline T_{8})^{k}_{l}  \overline
	{D}^{i} \overline{D}^{m} (H_{\overline6})^{il}_{m}
	+a_2 \Omega_{ccc}^{++} \varepsilon_{ijk} (\overline T_{8})^{k}_{l}  \overline{D}^{l} \overline{D}^{m} (H_{\overline6})^{ij}_{m} \nn\\
	&&+a_3  \Omega_{ccc}^{++} \varepsilon_{ijk} (\overline T_{8})^{k}_{l}  \overline{D}^{i} \overline{D}^{m} (H_{15})^{il}_{m} + a_4  \Omega_{ccc}^{++} (\overline T_{10})_{ijk} \overline{D}^{k} \overline{D}^{l} (H_{15})^{ij}_{l} \,.
\end{eqnarray}
Where the $a_i$'s are SU(3) irreducible nonperturbative amplitudes. The first three terms in Eq.(\ref{cccA}) denote the light baryon containing in the final states belongs to SU(3) octet, the last term denotes the light baryon in SU(3) decuplet. Feynman diagrams for these modes are given in Fig.~\ref{1.1DDT810}. Decay amplitudes for various channels can be deduced from the Hamiltonian in Eq.(\ref{cccA}), and are collected in Table~\ref{tab:ccc_A} (light baryon in octet) and Table~\ref{tab:ccc_B} (light baryon in decuplet).

\begin{table}
\caption{Amplitudes for $\Omega_{ccc}^{++}$ decays into two $D$-mesons and a light baryon (octet).}\label{tab:ccc_A}\begin{tabular}{cccc}\hline\hline
Channel & Amplitude & Channel & Amplitude \\\hline
$\Omega_{ccc}^{++}\to D^0  D^+  \Sigma^+ $ & $ -a_1+2 a_2-a_3$ &
$\Omega_{ccc}^{++}\to D^+  D^+  \Lambda^0 $ & $ \sqrt{\frac{2}{3}}(-a_1+2 a_2+3 a_3)$\\
$\Omega_{ccc}^{++}\to D^+  D^+  \Sigma^0 $ & $ \sqrt{2}(a_1-2 a_2+a_3)$ &
$\Omega_{ccc}^{++}\to D^+  D^+_s  \Xi^0 $ & $ -a_1+2 a_2+a_3$\\
\hline\hline
$\Omega_{ccc}^{++}\to D^0  D^+  {p} $ & $ \left(a_1-2 a_2+a_3\right) (-\sin\theta_{c})$ &
$\Omega_{ccc}^{++}\to D^0  D^+_s  \Sigma^+ $ & $ \left(a_1-2 a_2+a_3\right) (-\sin\theta_{c})$\\
$\Omega_{ccc}^{++}\to D^+  D^+  {n} $ & $ 2\left(-a_1+2 a_2+a_3\right) \sin\theta_{c}$ &
$\Omega_{ccc}^{++}\to D^+  D^+_s  \Lambda^0 $ & $ \frac{\left(a_1-2 a_2+3 a_3\right) }{\sqrt{6}}\sin\theta_{c}$\\
$\Omega_{ccc}^{++}\to D^+  D^+_s  \Sigma^0 $ & $ \frac{\left(a_1-2 a_2+3 a_3\right) }{\sqrt{2}}\sin\theta_{c}$ &
$\Omega_{ccc}^{++}\to D^+_s  D^+_s  \Xi^0 $ & $ 2\left(-a_1+2 a_2+a_3\right) \sin\theta_{c}$\\\hline\hline
$\Omega_{ccc}^{++}\to D^0  D^+_s  {p} $ & $ \left(a_1-2 a_2+a_3\right) \sin^2\theta_{c}$ &
$\Omega_{ccc}^{++}\to D^+  D^+_s  {n} $ & $ \left(a_1-2 a_2-a_3\right) \sin^2\theta_{c}$\\
$\Omega_{ccc}^{++}\to D^+_s  D^+_s  \Lambda^0 $ & $ -\sqrt{\frac{8}{3}} \left(a_1-2 a_2\right) \sin^2\theta_{c}$ &
$\Omega_{ccc}^{++}\to D^+_s  D^+_s  \Sigma^0 $ & $ -2\sqrt{2} a_3 \sin^2\theta_{c}$\\\hline
\hline
\end{tabular}
\end{table}

\begin{table}
\caption{Amplitudes for $\Omega_{ccc}^{++}$ decays into two $D$-mesons and a light baryon (decuplet).}\label{tab:ccc_B}\begin{tabular}{cccc}\hline\hline
Channel & Amplitude & Channel & Amplitude\\\hline
$\Omega_{ccc}^{++}\to D^0  D^+  \Sigma^{\prime+} $ & $ \frac{2 a_4}{\sqrt{3}}$ &
$\Omega_{ccc}^{++}\to D^+  D^+  \Sigma^{\prime0} $ & $ \sqrt{\frac{8}{3}} a_4$\\
$\Omega_{ccc}^{++}\to D^+_s  D^+  \Xi^{\prime0} $ & $ \frac{2 a_4}{\sqrt{3}}$\\\hline\hline
$\Omega_{ccc}^{++}\to D^0  D^+  \Delta^{+} $ & $ -\frac{2 a_4 }{\sqrt{3}}\sin\theta_{c}$ &
$\Omega_{ccc}^{++}\to D^0  D^+_s  \Sigma^{\prime+} $ & $ \frac{2 a_4 }{\sqrt{3}}\sin\theta_{c}$\\
$\Omega_{ccc}^{++}\to D^+  D^+  \Delta^{0} $ & $ -\frac{4 a_4 }{\sqrt{3}}\sin\theta_{c}$ &
$\Omega_{ccc}^{++}\to D^+_s  D^+_s  \Xi^{\prime0} $ & $ \frac{4 a_4 }{\sqrt{3}}\sin\theta_{c}$\\\hline\hline
$\Omega_{ccc}^{++}\to D^0  D^+_s  \Delta^{+} $ & $ \frac{2 a_4 }{\sqrt{3}}\sin^2\theta_{c}$ &
$\Omega_{ccc}^{++}\to D^+_s  D^+  \Delta^{0} $ & $ \frac{2 a_4 }{\sqrt{3}}\sin^2\theta_{c}$\\
$\Omega_{ccc}^{++}\to D^+_s  D^+_s  \Sigma^{\prime0} $ & $ \sqrt{\frac{8}{3}} a_4 \sin^2\theta_{c}$\\\hline
\hline
\end{tabular}
\end{table}

A few remarks are given in order:
\begin{enumerate}
  \item The initial state $\Omega_{ccc}^{++}$ is flavor SU(3) singlet, thus has no index to connect to the final states and $H_{\overline6, 15}$, this is a unique property comparing with previous works on weak decays of singly and doubly heavy baryons.
  \item Table~\ref{tab:ccc_A} and Table~\ref{tab:ccc_B} are arranged according to the decay amplitude's dependence on $\sin\theta_{c}$, $c\to s$ transition is proportional to $V_{cs}\sim 1$, while $c\to d$ transition has a Cabibbo suppressed CKM matrix element $V_{cs}\sim 0.2$.
  \item A number of relations for decay widths can be readily deduced from Table~\ref{tab:ccc_A} and Table~\ref{tab:ccc_B},
      \begin{eqnarray}
	      &&\Gamma(\Omega_{ccc}^{++}\to D^0D^+_s\Sigma^+)= { }\Gamma(\Omega_{ccc}^{++}\to D^+D^0{p})\,,\nn\\
	      &&\Gamma(\Omega_{ccc}^{++}\to D^+D^+{n})= { }\Gamma(\Omega_{ccc}^{++}\to D^+_sD^+_s\Xi^0)\,,\nn\\
          &&\Gamma(\Omega_{ccc}^{++}\to D^0D^+\Sigma^+)= \Gamma(\Omega_{ccc}^{++}\to D^+D^+\Sigma^0)\,,\nn\\
          &&\Gamma(\Omega_{ccc}^{++}\to D^+D^+_s\Lambda^0)= \frac{1}{3}\Gamma(\Omega_{ccc}^{++}\to D^+_sD^+\Sigma^0) \,, \nn\\
	      &&\Gamma(\Omega_{ccc}^{++}\to D^0  D^+  \Sigma^{\prime+})= { }\Gamma(\Omega_{ccc}^{++}\to D^+_s  D^+  \Xi^{\prime0})=\Gamma(\Omega_{ccc}^{++}\to D^+  D^+  \Sigma^{\prime0}) \,,\nn\\
	      &&\Gamma(\Omega_{ccc}^{++}\to D^0  D^+_s  \Delta^{+})= { }\Gamma(\Omega_{ccc}^{++}\to D^+_s  D^+  \Delta^{0} )=\Gamma(\Omega_{ccc}^{++}\to D^+_s  D^+_s  \Sigma^{\prime0}) \,,\nn\\
	      &&\Gamma(\Omega_{ccc}^{++}\to D^0  D^+  \Delta^{+} )= { }\Gamma(\Omega_{ccc}^{++}\to D^0  D^+_s  \Sigma^{\prime+}) \nn\\
          &&=\frac{1}{2}\Gamma(\Omega_{ccc}^{++}\to D^+  D^+  \Delta^{0})=\frac{1}{2}\Gamma(\Omega_{ccc}^{++}\to D^+_s  D^+_s  \Xi^{\prime0}) \,.
\end{eqnarray}
\end{enumerate}
However, it is necessary to point out that the above relationships between decay widths are obtained in the flavor SU(3) symmetry limit, in which the mass differences between final state hadrons have been ignored. Although the influence of identical particles on phase space integration has been considered, these relationships will be modified when calculating the kinematic corrections. Once the mass of $\Omega_{ccc}^{++}$ is experimentally measured in the future, a rigorous analysis would be necessary.

Particular decay processes of $\Omega_{ccc}^{++}$ in the detectors can be used as signatures to reconstruct this triply heavy baryon. The ground states of triply heavy baryons can decay only through the weak interaction. A interesting decay mode has been proposed by Ref.~\cite{Chen:2011mb}:
\begin{eqnarray}
  \Omega_{ccc}^{++} \,\,\red{\to}\,\, \Omega_{ccs}^{+}+\pi^+ \,\,\red{\to}\,\, \Omega_{css}^{0}+2\pi^+ \,\,\red{\to}\,\, \Omega_{sss}^{-}+3\pi^+ \,.
\end{eqnarray}
With this cascade mode, $\Omega_{ccc}^{++}$ finally decays to $\Omega_{ccc}^{++} + 3\pi^+$ and every step is Cabibbo-allowed. Having the results of previous works at hand~\cite{Geng:2017mxn,Wang:2017azm,Wang:2018utj}, we can write down other Cabibbo-allowed cascade modes of $\Omega_{ccc}^{++}$ which might be useful for finding this triply heavy baryon. They are collected in Table~\ref{cascade}

\begin{table}
\begin{scriptsize}
\caption{Cabibbo-allowed cascade modes of $\Omega_{ccc}^{++}$.}\label{cascade}\begin{tabular}{cccc||cccc}\hline\hline
   & Cascade & Channel & & & Cascade & Channel &\\\hline
   $\Omega_{ccc}^{++} \red{\to}$ & $\Omega_{ccs}^{+}+\pi^+ \red{\to}$  & $\Omega_{css}^{0}+2\pi^+ \red{\to}$  & $\Omega_{sss}^{-}+3\pi^+$
   & $\Omega_{ccc}^{++} \red{\to}$ & $\Xi_{cc}^{++}+\overline{K}^0 \red{\to}$  & $\Sigma_{c}^{++}+2\overline{K}^0 \red{\to}$  & $\Delta^{++}+3\overline{K}^0$ \\
   $\Omega_{ccc}^{++} \red{\to}$ & $\Omega_{ccs}^{+}+\pi^+ \red{\to}$  & $\Omega_{css}^{0}+2\pi^+ \red{\to}$  & $\Xi'^{0} + 2\pi^+ + \overline{K}^0$
   & $\Omega_{ccc}^{++} \red{\to}$ & $\Xi_{cc}^{++}+\overline{K}^0 \red{\to}$  & $\Sigma_{c}^{++}+2\overline{K}^0 \red{\to}$  & $\Sigma'^{+}+2\overline{K}^0+\pi^+$ \\
   $\Omega_{ccc}^{++} \red{\to}$ & $\Omega_{ccs}^{+}+\pi^+ \red{\to}$  & $\Omega_{css}^{0}+2\pi^+ \red{\to}$  & $\Xi^{0} + 2\pi^+ + \overline{K}^0$
   & $\Omega_{ccc}^{++} \red{\to}$ & $\Xi_{cc}^{++}+\overline{K}^0 \red{\to}$  & $\Sigma_{c}^{++}+2\overline{K}^0 \red{\to}$  & $\Sigma^{+}+2\overline{K}^0+\pi^+$ \\
   $\Omega_{ccc}^{++} \red{\to}\,\,$ & $\Omega_{ccs}^{+}+\pi^+ \red{\to}$  & $\Xi_{c}'^{+}/\Xi_{c}^{+} + \pi^+ + \overline{K}^0 \red{\to}$  & $\Sigma'^{+} + \pi^+ + 2\overline{K}^0$
   & $\Omega_{ccc}^{++} \red{\to}$ & $\Xi_{cc}^{++}+\overline{K}^0 \red{\to}$  & $\Xi_{c}'^{+}/\Xi_{c}^{+} + \overline{K}^0 +\pi^+ \red{\to}$  & $\Sigma'^{+} + \pi^+ + 2\overline{K}^0$ \\
   $\Omega_{ccc}^{++} \red{\to}\,\,$ & $\Omega_{ccs}^{+}+\pi^+ \red{\to}$  & $\Xi_{c}'^{+}/\Xi_{c}^{+} + \pi^+ + \overline{K}^0 \red{\to}$  & $\Xi'^{0} + 2\pi^+ + \overline{K}^0$
   & $\Omega_{ccc}^{++} \red{\to}$ & $\Xi_{cc}^{++}+\overline{K}^0 \red{\to}$  & $\Xi_{c}'^{+}/\Xi_{c}^{+} + \overline{K}^0 +\pi^+ \red{\to}$  & $\Xi'^{0} + 2\pi^+ + \overline{K}^0$ \\
   $\Omega_{ccc}^{++} \red{\to}\,\,$ & $\Omega_{ccs}^{+}+\pi^+ \red{\to}$  & $\Xi_{c}'^{+}/\Xi_{c}^{+} + \pi^+ + \overline{K}^0 \red{\to}$  & $\Sigma^{+} + \pi^+ + 2\overline{K}^0$
   & $\Omega_{ccc}^{++} \red{\to}$ & $\Xi_{cc}^{++}+\overline{K}^0 \red{\to}$  & $\Xi_{c}'^{+}/\Xi_{c}^{+} + \overline{K}^0 +\pi^+ \red{\to}$  & $\Sigma^{+} + \pi^+ + 2\overline{K}^0$ \\
   $\Omega_{ccc}^{++} \red{\to}\,\,$ & $\Omega_{ccs}^{+}+\pi^+ \red{\to}$  & $\Xi_{c}'^{+}/\Xi_{c}^{+} + \pi^+ + \overline{K}^0 \red{\to}$  & $\Xi^{0} + 2\pi^+ + \overline{K}^0$
   & $\Omega_{ccc}^{++} \red{\to}$ & $\Xi_{cc}^{++}+\overline{K}^0 \red{\to}$  & $\Xi_{c}'^{+}/\Xi_{c}^{+} + \overline{K}^0 +\pi^+ \red{\to}$  & $\Xi^{0} + 2\pi^+ + \overline{K}^0$ \\
\hline\hline
\end{tabular}
\end{scriptsize}
\end{table}

\subsection{Nonleptonic $\Omega_{bbb}^{-}$ decays}
For the bottom-quark decay, we can categorize the quark-level transitions into four kinds,
\begin{eqnarray}
\begin{aligned}
b \rightarrow c \bar{c} d / s \,, & \qquad b \rightarrow c \bar{u} d / s \,, \\
b \rightarrow u \bar{c} d / s \,, & \qquad b \rightarrow q \bar{q} q \,.
\end{aligned}
\end{eqnarray}
We will study $\Omega_{bbb}^{-}\to T_{bc} \, B_c $ and $\Omega_{bbb}^{-}\to T_{bc} \, B_c \, M$ for the first quark-level transition $b \rightarrow c \bar{c} d / s$ case. For the second $b \rightarrow c \bar{u} d / s$ case, decay modes $\Omega_{bbb}^{-}\to T_{bc} \, B $, $\Omega_{bbb}^{-}\to T_{bc} \, B \, M$ and $\Omega_{bbb}^{-}\to B \, B \, T_{c\bar3/6}$ will be discussed. The $b \rightarrow u \bar{c} d / s$ is highly suppressed by the corresponding CKM matrix elements as illustrated by Eq.(\ref{CKMNum}), therefore is not considered in this paper. For the last $b \rightarrow q \bar{q} q$ case, the decay modes $\Omega_{bbb}^{-}\to B \, B \, T_{8/10}$ would be analyzed .

\begin{figure}
\includegraphics[width=0.47\columnwidth]{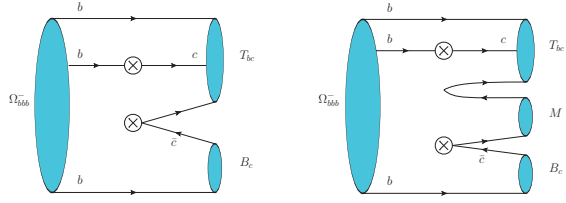}
\caption{Feynman diagrams for $\Omega_{bbb}^{-}$ decays into a doubly heavy baryon $T_{bc}$, plus a $B_c$ meson and a light meson.}
\label{2.2TbcBcM}
\end{figure}

The transition operator $b \rightarrow c \bar{c} d / s$ can form an SU(3) triplet with $(H_3)_2=V_{cd}^*$ and $(H_3)_3=V_{cs}^*$, one has the Hamiltonian for $\Omega_{bbb}^{-}$ decays into a doubly heavy baryon $T_{bc}$, plus a $B_c$ and a light meson:
\begin{eqnarray}\label{bbb1}
	{\cal H}_{\textit{eff}}&=& b_1  \Omega_{bbb}^{-} (\overline T_{bc})_{i}  B_{c}  (H_{3})^{i}
	+b_2  \Omega_{bbb}^{-} (\overline T_{bc})_{i}  B_{c} M^{i}_{j} (H_{3})^{j}\,.
\end{eqnarray}
The relevant Feynman diagrams are presented in Fig.~\ref{2.2TbcBcM}. Decay amplitudes for different channels are obtained by expanding the above Hamiltonian and are collected in Table~\ref{tab:bbb2}. These lead to the relations for decay widths:
\begin{eqnarray}	
  &&\Gamma(\Omega_{bbb}^{-}\to \Xi_{bc}^{0}\pi^0 B_c)= \frac{1}{2}\Gamma(\Omega_{bbb}^{-}\to \Xi_{bc}^{+}\pi^- B_c) \nn\\\
  &&= \frac{1}{2}\Gamma(\Omega_{bbb}^{-}\to \Omega_{bc}^{0}K^0 B_c) = 3\Gamma(\Omega_{bbb}^{-}\to \Xi_{bc}^{0}\eta B_c) \,.
\end{eqnarray}
\begin{table}
\caption{Amplitudes for $\Omega_{bbb}$ decays into $T_{bc}$ and a $B_c$ meson plus a light meson.}\label{tab:bbb2}\begin{tabular}{cccc}\hline\hline
Channel & Amplitude & Channel & Amplitude\\\hline
$\Omega_{bbb}^{-}\to \Xi_{bc}^{0}B_c $ & $ b_1 V_{cb} V_{cs}^*$\\
$\Omega_{bbb}^{-}\to \Xi_{bc}^{+}  \pi^-  B_c $ & $ b_2 V_{cb} V_{cs}^*$ &
$\Omega_{bbb}^{-}\to \Xi_{bc}^{0}  \pi^0  B_c $ & $ -\frac{b_2 V_{cb} V_{cs}^*}{\sqrt{2}}$\\
$\Omega_{bbb}^{-}\to \Xi_{bc}^{0}  \eta  B_c $ & $ \frac{b_2 V_{cb} V_{cs}^*}{\sqrt{6}}$ &
$\Omega_{bbb}^{-}\to \Omega_{bc}^{0}  K^0  B_c $ & $ b_2 V_{cb} V_{cs}^*$\\\hline
\hline
\end{tabular}
\end{table}

\begin{figure}
\includegraphics[width=1\columnwidth]{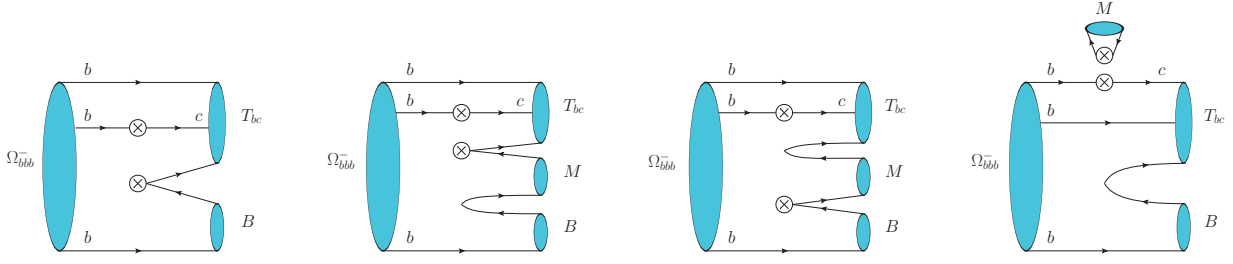}
\caption{Feynman diagrams for $\Omega_{bbb}^{-}$ decays into a doubly heavy baryon $T_{bc}$, plus a $B$-meson and a light meson.}
\label{2.1TbcBM}
\end{figure}

The transition operator $b \rightarrow c \bar{u} d / s$ forms an SU(3) octet $H_{8}$ with nonzero entries
\begin{eqnarray}\label{Hfor2type}
 \left(H_{8}\right)_{1}^{2}=V_{c b} V_{u d}^{*} \,, \quad \left(H_{8}\right)_{1}^{3}=V_{c b} V_{u s}^{*} \,.
\end{eqnarray}
Thus, we have the effective Hamiltonian for $\Omega_{bbb}^{-}$ decays into a doubly heavy baryon $T_{bc}$, plus a $B$-meson and a light meson:
\begin{eqnarray}\label{bbb2}
   	{\cal H}_{\textit{eff}}&=& c_1 \Omega_{bbb}^{-}  (\overline T_{bc})_{i}  \overline{B}^{j}  (H_{8})^{i}_{j}
   	+c_2  \Omega_{bbb}^{-}  (\overline T_{bc})_{i}  \overline{B}^{i} M^{j}_{k} (H_{8})^{k}_{j}\nn\\
   	&&+c_3  \Omega_{bbb}^{-}  (\overline T_{bc})_{i}  \overline{B}^{j} M^{i}_{k} (H_{8})^{k}_{j}
    +c_4  \Omega_{bbb}^{-}  (\overline T_{bc})_{i}  \overline{B}^{j} M^{k}_{j} (H_{8})^{i}_{k}.
\end{eqnarray}
The Feynman diagrams for these decays modes are given in Fig.~\ref{2.1TbcBM}. Expanding the above equations, we will obtain the decay amplitudes given in Table~\ref{tab:bbb3}, which lead to one relation for decay widths:
\begin{eqnarray}
    \Gamma(\Omega_{bbb}^{-}\to \Omega_{bc}^{0}B^-\pi^0 )= \frac{1}{2}\Gamma(\Omega_{bbb}^{-}\to \Omega_{bc}^{0}\overline B^0\pi^- ) \,.
 \end{eqnarray}

\begin{table}
\caption{Amplitudes for $\Omega_{bbb}^{-}$ decays into a $T_{bc}$ and a $B$-meson and a light meson.}\label{tab:bbb3}\begin{tabular}{cccc}\hline\hline
Channel & Amplitude & Channel & Amplitude \\\hline
$\Omega_{bbb}^{-}\to \Xi_{bc}^{0}  B^- $ & $ c_1 V_{cb} V_{ud}^*$ & $\Omega_{bbb}^{-}\to \Omega_{bc}^{0}  B^-  $ & $ c_1 V_{cb} V_{us}^* $\\
$\Omega_{bbb}^{-}\to \Xi_{bc}^{+}  B^-  \pi^-  $ & $ \left(c_2+c_3\right) V_{cb} V_{ud}^*$ &
$\Omega_{bbb}^{-}\to \Xi_{bc}^{+}  B^-  K^-  $ & $ \left(c_2+c_3\right) V_{cb} V_{us}^*$\\
$\Omega_{bbb}^{-}\to \Xi_{bc}^{0}  B^-  \pi^0  $ & $ -\frac{\left(c_3-c_4\right) V_{cb} V_{ud}^*}{\sqrt{2}}$ &
$\Omega_{bbb}^{-}\to \Xi_{bc}^{0}  B^-  \overline K^0  $ & $ c_3 V_{cb} V_{us}^*$\\
$\Omega_{bbb}^{-}\to \Xi_{bc}^{0}  B^-  \eta  $ & $ \frac{\left(c_3+c_4\right) V_{cb} V_{ud}^*}{\sqrt{6}}$ &
$\Omega_{bbb}^{-}\to \Xi_{bc}^{0}  \overline B^0  \pi^-  $ & $ \left(c_2+c_4\right) V_{cb} V_{ud}^*$\\
$\Omega_{bbb}^{-}\to \Xi_{bc}^{0}  \overline B^0  K^-  $ & $ c_2 V_{cb} V_{us}^*$ &
$\Omega_{bbb}^{-}\to \Xi_{bc}^{0}  \overline B^0_s  K^-  $ & $ c_4 V_{cb} V_{ud}^*$\\
$\Omega_{bbb}^{-}\to \Omega_{bc}^{0}  B^-  \pi^0  $ & $ \frac{c_4 V_{cb} V_{us}^*}{\sqrt{2}}$ &
$\Omega_{bbb}^{-}\to \Omega_{bc}^{0}  B^-  K^0  $ & $ c_3 V_{cb} V_{ud}^*$\\
$\Omega_{bbb}^{-}\to \Omega_{bc}^{0}  B^-  \eta  $ & $ \frac{\left(c_4-2 c_3\right) V_{cb} V_{us}^*}{\sqrt{6}}$ &
$\Omega_{bbb}^{-}\to \Omega_{bc}^{0}  \overline B^0  \pi^-  $ & $ c_4 V_{cb} V_{us}^*$\\
$\Omega_{bbb}^{-}\to \Omega_{bc}^{0}  \overline B^0_s  \pi^-  $ & $ c_2 V_{cb} V_{ud}^*$ &
$\Omega_{bbb}^{-}\to \Omega_{bc}^{0}  \overline B^0_s  K^-  $ & $ \left(c_2+c_4\right) V_{cb} V_{us}^*$\\\hline
\hline
\end{tabular}
\end{table}

\begin{figure}
\includegraphics[width=0.25\columnwidth]{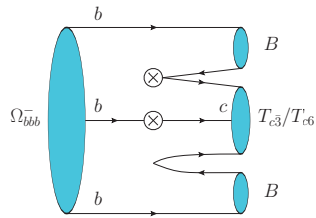}
\caption{Feynman diagrams for $\Omega_{bbb}^{-}$ decays into two $B$-mesons and a singly charmed baryon (antitriplet or sextet).}
\label{2.5BBTc3c6}
\end{figure}

The quark-level transition operator for $\Omega_{bbb}^{-}$ decays into two $B$-mesons and one singly charmed baryon (antitriplet or sextex) is also $b \rightarrow c \bar{u} d / s$ whose nonzero entries have been already shown in Eq.(\ref{Hfor2type}). The effective Hamiltonian is derived as
\begin{eqnarray}\label{H:1}
	{\cal H}_{\textit{eff}}&=& d_1  \Omega_{bbb}^{-} \overline{B}^{i} \overline{B}^{j} (\overline{T}_{c\overline3})_{[ik]}   (H_{8})^{k}_{j}
	+d_2 \Omega_{bbb}^{-} \overline{B}^{i} \overline{B}^{j} (\overline{T}_{c6})_{[ik]}   (H_{8})^{k}_{j} \,.
\end{eqnarray}
The Feynman diagrams for these decays channels are given in Fig.~\ref{2.5BBTc3c6} and the decay amplitudes are collected in Table~\ref{tab:bbb4}, the relations of different decay channels are presented below:
\begin{eqnarray}
    &&\Gamma(\Omega_{bbb}^{-}\to B^-B^-\Lambda_c^+)= 2\Gamma(\Omega_{bbb}^{-}\to B^-\overline B^0_s\Xi_c^0) \,,\nn\\
    &&\Gamma(\Omega_{bbb}^{-}\to B^-B^-\Xi_c^+)= 2\Gamma(\Omega_{bbb}^{-}\to B^-\overline B^0\Xi_c^0) \,,\nn\\
    &&\Gamma(\Omega_{bbb}^{-}\to B^-B^-\Sigma_{c}^{+})= 2\Gamma(\Omega_{bbb}^{-}\to \overline B^0_sB^-\Xi_{c}^{\prime0})= \Gamma(\Omega_{bbb}^{-}\to B^-\overline B^0\Sigma_{c}^{0}) \,,\nn\\
    &&\Gamma(\Omega_{bbb}^{-}\to B^-B^-\Xi_{c}^{\prime+})= 2\Gamma(\Omega_{bbb}^{-}\to B^-\overline B^0\Xi_{c}^{\prime0})= \Gamma(\Omega_{bbb}^{-}\to B^-\overline B^0_s\Omega_{c}^{0}) \,.
 \end{eqnarray}

\begin{table}
\caption{Amplitudes for $\Omega_{bbb}^{-}$ decays into two $B$-mesons and a singly charmed baryon (antitriplet or sextet).}\label{tab:bbb4}\begin{tabular}{cccc}\hline\hline
	Channel & Amplitude & Channel & Amplitude \\\hline
$\Omega_{bbb}^{-}\to B^-  B^-  \Lambda_c^+ $ & $ 2 d_1 V_{cb} V_{ud}^*$ &
$\Omega_{bbb}^{-}\to B^-  B^-  \Xi_c^+ $ & $ 2 d_1 V_{cb} V_{us}^*$\\
$\Omega_{bbb}^{-}\to B^-  \overline B^0  \Xi_c^0 $ & $ d_1 V_{cb} V_{us}^*$ &
$\Omega_{bbb}^{-}\to B^-  \overline B^0_s  \Xi_c^0 $ & $ -d_1 V_{cb} V_{ud}^*$\\\hline\hline
$\Omega_{bbb}^{-}\to B^-  B^-  \Sigma_{c}^{+} $ & $ \sqrt{2} d_2 V_{cb} V_{ud}^*$ &
$\Omega_{bbb}^{-}\to B^-  B^-  \Xi_{c}^{\prime+} $ & $ \sqrt{2} d_2 V_{cb} V_{us}^*$\\
$\Omega_{bbb}^{-}\to B^-  \overline B^0  \Sigma_{c}^{0} $ & $ d_2 V_{cb} V_{ud}^*$ &
$\Omega_{bbb}^{-}\to B^-  \overline B^0  \Xi_{c}^{\prime0} $ & $ \frac{d_2 V_{cb} V_{us}^*}{\sqrt{2}}$\\
$\Omega_{bbb}^{-}\to B^-  \overline B^0_s  \Xi_{c}^{\prime0} $ & $ \frac{d_2 V_{cb} V_{ud}^*}{\sqrt{2}}$ &
$\Omega_{bbb}^{-}\to B^-  \overline B^0_s  \Omega_{c}^{0} $ & $ d_2 V_{cb} V_{us}^*$\\\hline
\hline
\end{tabular}
\end{table}

For the last kind of quark-level transition $b \rightarrow q \bar{q} q$, we will study the decay modes $\Omega_{bbb}^{-}\to B \, B \, T_{8/10}$. The charmless $b\to q/s$ transition is depicted by the weak Hamiltonian $\cal H_{\textit{e.w.}}$,
\begin{eqnarray}\label{HincludePenguin}
\mathcal{H}_{\text {e.w. }}=\frac{G_{F}}{\sqrt{2}}\left\{V_{u b} V_{u q}^{*}\left[C_{1} O_{1}^{\bar{u} u}+C_{2} O_{2}^{\bar{u} u}\right]-V_{t b} V_{t q}^{*}\left[\sum_{i=3}^{10} C_{i} O_{i}\right]\right\}+\text { H.c. } \,,
\end{eqnarray}
where the explicit expressions for $O_i$ can be found in Ref.~\cite{Buras:1998raa}. Penguin operators in Eq.(\ref{HincludePenguin}) are not suppressed, they behave as 3 representation in the SU(3) group,
\begin{eqnarray}
  \left(H_3\right)^2=1 ~(\textrm{for}~ \Delta S=0, i.e., b\to d ~\textrm{case}) \,, \qquad \left(H_3\right)^3=1 ~(\textrm{for}~ \Delta S=1, i.e., b\to s ~\textrm{case}) \,.
\end{eqnarray}
Tree operators can be decomposed in terms of a vector $H_{3}$, two traceless tensors $H_{\overline6}$ and $H_{15}$ whose nonzero entries are
\begin{eqnarray}
\left\{\begin{array}{l} \left(H_{\overline{6}}\right)_{1}^{12}=-\left(H_{\overline{6}}\right)_{1}^{21}=\left(H_{\overline{6}}\right)_{3}^{23}=-\left(H_{\overline{6}}\right)_{3}^{32}=1 \,, \\
2\left(H_{15}\right)_{1}^{12} =2\left(H_{15}\right)_{1}^{21}=-3\left(H_{15}\right)_{2}^{22} \\
=-6\left(H_{15}\right)_{3}^{23}=-6\left(H_{15}\right)_{3}^{32}=6 \,. \end{array}\right. \qquad (\textrm{for}~ \Delta S=0, i.e., b\to d ~\textrm{case})\\
\left\{\begin{array}{l} \left(H_{\overline{6}}\right)_{1}^{13}=-\left(H_{\overline{6}}\right)_{1}^{31}=\left(H_{\overline{6}}\right)_{2}^{32}=-\left(H_{\overline{6}}\right)_{2}^{23}=1 \,, \\
2\left(H_{15}\right)_{1}^{13} =2\left(H_{15}\right)_{1}^{31}=-3\left(H_{15}\right)_{3}^{33} \\
=-6\left(H_{15}\right)_{2}^{32}=-6\left(H_{15}\right)_{2}^{23}=6 \,. \end{array}\right. \qquad (\textrm{for}~ \Delta S=1, i.e., b\to s ~\textrm{case})
\end{eqnarray}

\begin{figure}
\includegraphics[width=0.50\columnwidth]{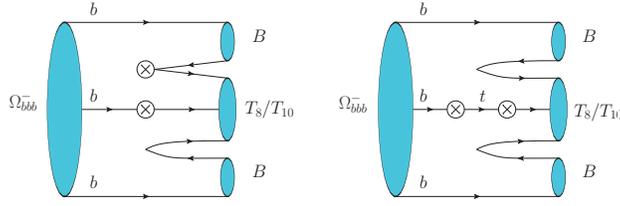}
\caption{Feynman diagrams for $\Omega_{bbb}^{-}$ decays into two $B$-mesons and a light baryon (octet or decuplet).}
\label{2.3BBT810}
\end{figure}

Thus, for $\Omega_{bbb}^{-}$ decays into two $B$-mesons and one light baryon (octet), the effective Hamiltonian is given as
\begin{eqnarray}
	{\cal H}_{\textit{eff}}&=& f_1  \Omega_{bbb}^{-}  \overline B^{i}  \overline{B}^{j} \varepsilon_{ijk} (\overline{T}_{8})^{k}_{l} (H_{3})^{l}
	+f_2  \Omega_{bbb}^{-}  \overline B^{i}  \overline{B}^{l} \varepsilon_{ijk} (\overline{T}_{8})^{k}_{l} (H_{3})^{j} \nn\\
	&&+f_3 \Omega_{bbb}^{-}  \overline B^{i}  \overline{B}^{m} \varepsilon_{ijk} (\overline{T}_{8})^{k}_{l} (H_{\overline6})^{jl}_{m}
	+f_4  \Omega_{bbb}^{-}  \overline B^{m}  \overline{B}^{l} \varepsilon_{ijk} (\overline{T}_{8})^{k}_{l} (H_{\overline6})^{ij}_{m} \nn\\
	&&+f_5  \Omega_{bbb}^{-}  \overline B^{i}  \overline{B}^{m} \varepsilon_{ijk} (\overline{T}_{8})^{k}_{l} (H_{15})^{jl}_{m} \,.
\end{eqnarray}
Similarly, the decay amplitudes are obtained and collected in Table~\ref{tab:bbb5}, the corresponding Feynman diagrams are presented in Fig.~\ref{2.3BBT810}. Two relations for decay widths can be read off
\begin{eqnarray}
    \Gamma(\Omega_{bbb}^{-}\to \overline B^0\overline B^0_s\Sigma^-)= \frac{1}{2}\Gamma(\Omega_{bbb}^{-}\to \overline B^0_s\overline B^0_s\Xi^-) \,,\nn\\
    \Gamma(\Omega_{bbb}^{-}\to \overline B^0\overline B^0\Sigma^-)= 2\Gamma(\Omega_{bbb}^{-}\to \overline B^0\overline B^0_s\Xi^-)\,.
\end{eqnarray}

\begin{table}
\caption{Amplitudes for $\Omega_{bbb}^{-}$ decays into two $B$-mesons and a light baryon (octet).}\label{tab:bbb5}\begin{tabular}{cccc}\hline\hline
Channel & Amplitude & Channel & Amplitude \\\hline
$\Omega_{bbb}^{-}\to B^-  B^-  {p} $ & $ 2(f_2-f_3+2 f_4+3 f_5)$ &
$\Omega_{bbb}^{-}\to B^-  \overline B^0  {n} $ & $ f_2-f_3+2 f_4-5 f_5$\\
$\Omega_{bbb}^{-}\to B^-  \overline B^0_s  \Lambda^0 $ & $ \frac{-3 f_2+f_3-2 f_4+3 f_5}{\sqrt{6}}$ &
$\Omega_{bbb}^{-}\to B^-  \overline B^0_s  \Sigma^0 $ & $ -\frac{f_2+f_3-2 f_4+7 f_5}{\sqrt{2}}$\\
$\Omega_{bbb}^{-}\to \overline B^0  \overline B^0_s  \Sigma^- $ & $ -f_2-f_3+2 f_4+f_5$ &
$\Omega_{bbb}^{-}\to \overline B^0_s  \overline B^0_s  \Xi^- $ & $ 2( -f_2-f_3+2 f_4+f_5)$\\\hline\hline
$\Omega_{bbb}^{-}\to B^-  B^-  \Sigma^+ $ & $ 2 (-f'_2+f'_3-2 f'_4-3 f'_5)$ &
$\Omega_{bbb}^{-}\to B^-  \overline B^0  \Lambda^0 $ & $ \sqrt{\frac{2}{3}} \left(f'_3-2 f'_4+6 f'_5\right)$\\
$\Omega_{bbb}^{-}\to B^-  \overline B^0  \Sigma^0 $ & $ \sqrt{2} \left(f'_2+f'_5\right)$ &
$\Omega_{bbb}^{-}\to B^-  \overline B^0_s  \Xi^0 $ & $ -f'_2+f'_3-2 f'_4+5 f'_5$\\
$\Omega_{bbb}^{-}\to \overline B^0  \overline B^0  \Sigma^- $ & $ 2 (f'_2+f'_3-2 f'_4-f'_5)$ &
$\Omega_{bbb}^{-}\to \overline B^0_s  \overline B^0  \Xi^- $ & $ f'_2+f'_3-2 f'_4-f'_5$\\\hline
\hline
\end{tabular}
\end{table}

For the case $\Omega_{bbb}^{-}$ decays into two $B$-mesons and one light baryon (decuplet), the effective Hamiltonian is given as
\begin{eqnarray}
	{\cal H}_{\textit{eff}}&=& g_1  \Omega_{bbb}^{-} \overline{B}^{i} \overline{B}^{j} (\overline T_{10})_{ijk}   (H_{3})^{k}
	+g_2 \Omega_{bbb}^{-} \overline{B}^{i} \overline{B}^{l} (\overline T_{10})_{ijk}   (H_{15})^{jk}_{l} \,.
\end{eqnarray}
The decay amplitudes are obtained and collected in Table~\ref{tab:bbb6}. Various relations for decay widths can be deduced:

\begin{table}
\caption{Amplitudes for $\Omega_{bbb}^{-}$ decays into two $B$-mesons and a light baryon (decuplet)}\label{tab:bbb6}\begin{tabular}{cccc}\hline\hline
	Channel & Amplitude & Channel & Amplitude \\\hline
$\Omega_{bbb}^{-}\to B^-  B^-  \Delta^{+} $ & $ \sqrt{\frac{4}{3}}(g_1+6 g_2)$ &
$\Omega_{bbb}^{-}\to B^-  \overline B^0  \Delta^{0} $ & $ \sqrt{\frac{4}{3}}\left(g_1+2 g_2\right)$\\
$\Omega_{bbb}^{-}\to B^-  \overline B^0_s  \Sigma^{\prime0} $ & $ \sqrt{\frac{2}{3}} \left(g_1+2 g_2\right)$ &
$\Omega_{bbb}^{-}\to \overline B^0  \overline B^0  \Delta^{-} $ & $ 2(g_1-2 g_2)$\\
$\Omega_{bbb}^{-}\to \overline B^0  \overline B^0_s  \Sigma^{\prime-} $ & $ \sqrt{\frac{4}{3}}\left(g_1-2 g_2\right)$ &
$\Omega_{bbb}^{-}\to \overline B^0_s  \overline B^0_s  \Xi^{\prime-} $ & $ \sqrt{\frac{4}{3}}(g_1-2 g_2)$\\\hline\hline
$\Omega_{bbb}^{-}\to B^-  B^-  \Sigma^{\prime+} $ & $ \sqrt{\frac{4}{3}}(g'_1+6 g'_2)$ &
$\Omega_{bbb}^{-}\to B^-  \overline B^0  \Sigma^{\prime0} $ & $ \sqrt{\frac{2}{3}} \left(g'_1+2 g'_2\right)$\\
$\Omega_{bbb}^{-}\to B^-  \overline B^0_s  \Xi^{\prime0} $ & $ \sqrt{\frac{4}{3}}\left(g'_1+2 g'_2\right)$ &
$\Omega_{bbb}^{-}\to \overline B^0  \overline B^0  \Sigma^{\prime-} $ & $ \sqrt{\frac{4}{3}}(g'_1-2 g'_2)$\\
$\Omega_{bbb}^{-}\to \overline B^0  \overline B^0_s  \Xi^{\prime-} $ & $ \sqrt{\frac{4}{3}}\left(g'_1-2 g'_2\right)$ &
$\Omega_{bbb}^{-}\to \overline B^0_s  B^-  \Xi^{\prime0} $ & $ \sqrt{\frac{4}{3}}\left(g'_1+2 g'_2\right)$\\
$\Omega_{bbb}^{-}\to \overline B^0_s  \overline B^0_s  \Omega^- $ & $ 2(g'_1-2 g'_2)$\\\hline
\hline
\end{tabular}
\end{table}
\begin{eqnarray}
    &&\Gamma(\Omega_{bbb}^{-}\to B^-\overline B^0\Delta^{0})= 2\Gamma(\Omega_{bbb}^{-}\to B^-\overline B^0_s\Sigma^{\prime0}) \,,\nn\\
    &&\Gamma(\Omega_{bbb}^{-}\to \overline B^0\overline B^0_s\Sigma^{\prime-})= 2\Gamma(\Omega_{bbb}^{-}\to \overline B^0_s\overline B^0_s\Xi^{\prime-}) \,,\nn\\
    &&\Gamma(\Omega_{bbb}^{-}\to \overline B^0B^-\Sigma^{\prime0})= \frac{1}{2}\Gamma(\Omega_{bbb}^{-}\to \overline B^0_sB^-\Xi^{\prime0}) \,,\nn\\
    &&\Gamma(\Omega_{bbb}^{-}\to \overline B^0\overline B^0_s\Xi^{\prime-})= \frac{2}{3}\Gamma(\Omega_{bbb}^{-}\to \overline B^0_s\overline B^0_s\Omega^-) \,,\nn\\
    &&\Gamma(\Omega_{bbb}^{-}\to \overline B^0\overline B^0\Delta^{-})= \frac{3}{2}\Gamma(\Omega_{bbb}^{-}\to \overline B^0_s\overline B^0\Sigma^{\prime-})= 3\Gamma(\Omega_{bbb}^{-}\to \overline B^0_s\overline B^0_s\Xi^{\prime-}) \,,\nn\\
    &&\Gamma(\Omega_{bbb}^{-}\to \overline B^0\overline B^0\Sigma^{\prime-})= \frac{1}{2}\Gamma(\Omega_{bbb}^{-}\to \overline B^0_s\overline B^0\Xi^{\prime-}) = \frac{1}{3}\Gamma(\Omega_{bbb}^{-}\to \overline B^0_s\overline B^0_s\Omega^-) \,.
\end{eqnarray}

A few remarks are given in order:
\begin{enumerate}
  \item The channels in Table~\ref{tab:bbb5} and Table~\ref{tab:bbb6} are arranged according to its quark level transition is whether $b\to d$ or $b\to s$. Note that their CKM matrix elements which have been absorbed in the nonperturbative coefficients are different, therefore the coefficients in SU(3) irreducible amplitudes for the $b\to s$ transition are primed ($f'_i$ and $g'_i$).
  \item One can infer that the typical branching fractions are at the order $10^{-6}$ through a simple analogy with the $B$-meson decay data. Thus there is little chance to discover the triply bottom baryon through these decay channels, but they can be utilized to study the direct CP asymmetries once large amount of data have been accumulated in future~\cite{Wang:2018utj}.
\end{enumerate}

\subsection{Nonleptonic $\Omega_{ccb}^{+}$ and $\Omega_{cbb}^{0}$ decays}
Most decay modes of $\Omega_{ccb}^{+}$ and $\Omega_{cbb}^{0}$ can be obtained from the results of $\Omega_{ccc}^{++}$ and $\Omega_{bbb}^{-}$ with some replacements. For example, decays of $\Omega_{ccb}^{+}$ induced by the charm quark can be obtained from the ones of $\Omega_{ccc}^{++}$ through replacing one charmed meson by the corresponding bottom meson, a charmed baryon by the corresponding bottom baryon, or a doubly charmed baryon $T_{cc}$ by its counterpart $T_{bc}$. In addition, there is another kind of decay modes of the mixed triply heavy baryons, i.e., the $W$-exchange transition which has been discussed in Ref.~\cite{Wang:2018utj}. Therefore we won't explicitly show the effective Hamiltonian for various decay modes of $\Omega_{ccb}^{+}$ and $\Omega_{cbb}^{0}$ here.

\begin{figure}
\includegraphics[width=0.47\columnwidth]{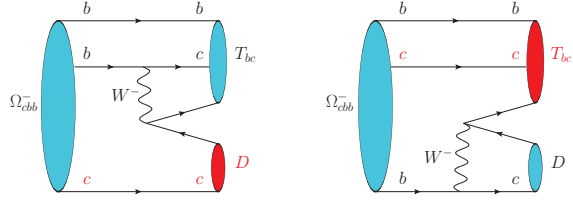}
\caption{Feynman diagrams for $\Omega_{cbb}^{0}$ decays into a doubly heavy baryon $T_{bc}$ and a $D$-meson. The final hadron into which the spectator $c$ quark flows is marked with red.}
\label{Omega_cbb}
\end{figure}

As we have mentioned before, one significant advantage of the SU(3) analysis is that it is independent of the factorization details, this can be clearly revealed through, for instance, the weak decay of $\Omega_{cbb}^{0}$ into a mixed doubly heavy baryon $T_{bc}$ and a $D$-meson. There are two possibilities in this weak decay: the spectator $c$ quark in $\Omega_{cbb}^{0}$ may flow in the final mixed doubly heavy baryon $T_{bc}$ or interact with a light antiquark to form a $D$-meson after hadronization. The typical Feynman diagrams corresponding to these two cases are depicted in Fig.~\ref{Omega_cbb}. However, from the perspective of flavor SU(3) analysis, there is no difference between these two cases since the heavy $c$ quark is SU(3) singlet. As long as one assumes that the flavor SU(3) symmetry is approximately preserved in the weak decays of triply heavy baryon, the SU(3) transformation invariant effective Hamiltonian can be written as ${\cal H}_{\textit{eff}} = h_1  \Omega_{cbb}^{0} (\overline T_{bc})_{i} \overline{D}^{j} (H_{8})^{i}_{j}$.

\section{Discussions and Conclusions}
\label{sec:conclusions}
Based on the above analysis in Sec.~\ref{sec:nonleptonic}, a collection of Cabibbo-allowed decay channels for $\Omega_{ccc}^{++}$ and CKM-allowed decay channels for $\Omega_{bbb}^{-}$ has been presented in Table~\ref{CKM_allowed}. For the $\Omega_{ccc}^{++}$ decay, the branching fractions for the Cabibbo-allowed processes can reach a few percent, thus might presumably lead to discovery of triply charmed baryon. For the $\Omega_{bbb}^{-}$ decay, the CKM-allowed largest branching fraction might reach $10^{-3}$, which would be even much smaller when considering detecting charmless final states in experiment.
	\begin{table}
		\caption{Cabibbo-allowed decay channels for $\Omega_{ccc}^{++}$ and CKM-allowed decay channels for $\Omega_{bbb}^{-}$}\label{CKM_allowed}\begin{tabular}{cccc}\hline\hline
			Channel & Channel & Channel & Channel\\\hline
			$\Omega_{ccc}^{++}\to D^0  D^+  \Sigma^+ $  \quad& $\Omega_{ccc}^{++}\to D^+  D^+  \Lambda^0 $ \quad &
			$\Omega_{ccc}^{++}\to D^+  D^+  \Sigma^0 $  \quad& $\Omega_{ccc}^{++}\to D^+  D^+_s  \Xi^0 $\\
			$\Omega_{ccc}^{++}\to D^0  D^+  \Sigma^{\prime+} $ \quad&
			$\Omega_{ccc}^{++}\to D^+  D^+  \Sigma^{\prime0} $  \quad&
			$\Omega_{ccc}^{++}\to D^+_s  D^+  \Xi^{\prime0} $ \\\hline\hline
            $\Omega_{bbb}^{-}\to \Xi_{bc}^{0}B_c $ \quad& $\Omega_{bbb}^{-}\to \Xi_{bc}^{0}  B^- $\\
			$\Omega_{bbb}^{-}\to \Omega_{bc}^{0}  K^0  B_c $ \quad&
			$\Omega_{bbb}^{-}\to \Xi_{bc}^{+}  \pi^-  B_c $  \quad&
			$\Omega_{bbb}^{-}\to \Xi_{bc}^{0}  \pi^0  B_c $  \quad& $\Omega_{bbb}^{-}\to \Xi_{bc}^{0}  \eta  B_c $ \\
			$\Omega_{bbb}^{-}\to \Omega_{bc}^{0}  \overline B^0_s  \pi^-  $ \quad& $\Omega_{bbb}^{-}\to \Xi_{bc}^{+}  B^-  \pi^-  $ \quad&
			$\Omega_{bbb}^{-}\to \Xi_{bc}^{0}  B^-  \pi^0  $  \quad& $\Omega_{bbb}^{-}\to \Xi_{bc}^{0}  B^-  \eta  $ \\
            $\Omega_{bbb}^{-}\to \Xi_{bc}^{0}  \overline B^0  \pi^-  $ \quad&
			$\Omega_{bbb}^{-}\to \Xi_{bc}^{0}  \overline B^0_s  K^-  $ \quad& $\Omega_{bbb}^{-}\to \Omega_{bc}^{0}  B^-  K^0  $ \quad&
			$\Omega_{bbb}^{-}\to B^-  B^-  \Lambda_c^+ $ \\
            $\Omega_{bbb}^{-}\to B^-  \overline B^0_s  \Xi_c^0 $ \quad& $\Omega_{bbb}^{-}\to B^-  B^-  \Sigma_{c}^{+} $ \quad& $\Omega_{bbb}^{-}\to B^-  \overline B^0  \Sigma_{c}^{0} $ \quad& $\Omega_{bbb}^{-}\to B^-  \overline B^0_s  \Xi_{c}^{\prime0} $
			\\\hline\hline
		\end{tabular}
	\end{table}

The triply heavy baryons are of considerable theoretical interests, since they are free of light quark contamination and can help to probe the interplay between perturbative and nonperturbative QCD~\cite{Brambilla:2013vx,Wei:2016jyk}. The observation of doubly heavy baryon makes it more reliable to look forward the triply heavy baryon in future colliders such as the high luminosity LHC.

This work is an extension of previous studies, we have systematically analyzed the nonleptonic weak decays of triply heavy baryons. Decay amplitudes for these processes have been parametrized in terms of the SU(3) irreducible nonperturbative amplitudes ($a_i$'s $\sim$ $g_i$'s). A number of relations for the partial decay widths can be deduced from these results and can be examined once we have a large amount of data in future.
We also list the decay channels of triply heavy baryons and some cascade decay modes of $\Omega_{ccc}^{++}$ which likely to be used to reconstructing in experiments. It is worth emphasizing here, that the triply heavy baryons are still absent in particle data booklet after the discovery of the heavy quarkonium $J/\psi$ over four decades. Therefore we encourage our experimental colleagues performing searches of this kind of particles since the reward could be high and will be a milestone in hadron physics.

\section*{Acknowledgements}
The authors are grateful to Professor Wei Wang and Dr. Ye Xing for inspiring discussions and valuable comments. When this work was about to finalize, Zhengzhou encountered unprecedented rainstorm, we want to express our gratitude to the people of Zhengzhou for their bravery in this extreme weather event. F.H is supported in part by Natural Science Foundation of China under grant Nos. 11735010, U2032102, 11653003, Natural Science Foundation of Shanghai under grant No. 15DZ2272100. J.X. is supported in part by National Natural Science Foundation of China under Grant No. 12047545.

\end{document}